\documentclass[prd,eqsecnum,twocolumn,amsfonts,amssymb,showpacs]{revtex4}

\usepackage{CJK}

\usepackage{graphicx}

\usepackage{bm}

\newcommand{\beq}{\begin{equation}}
\newcommand{\eeq}{\end{equation}}
\newcommand{\beqs}{\begin{eqnarray}}
\newcommand{\eeqs}{\end{eqnarray}}

\newcommand{\pslash}{p\hspace{-0.067in}\slash}

\begin{document}

\begin{CJK*}{UTF8}{}

\title{Radiative decays of $1^{++}$ heavy mesons in the covariant
light-front approach}

\author{Yan-Liang Shi (\CJKfamily{bsmi}石炎亮)}

\affiliation{C. N. Yang Institute for Theoretical Physics,
Stony Brook University, Stony Brook, N. Y. 11794 }

\begin{abstract}

  We calculate the predicted width for the radiative decay of a $1^{++}$ heavy
  meson via the channel $1^{++} \to 1^{--} +\gamma$ in the covariant
  light-front quark model.  Specifically, we compute the decay widths for
  $\chi_{c1}(1P) \to J/\psi + \gamma$ and $\chi_{b1}(nP) \to \Upsilon(n'S) +
  \gamma$. The results are compared with experimental data and with predictions
  from calculations based on nonrelativistic models and their extensions to
  include relativistic effects.

\end{abstract}


\maketitle

\end{CJK*}




\section{Introduction}

Experimental observations and theoretical studies of heavy quarkonium $Q \bar
Q$ states have played a very valuable role in elucidating the properties of
quantum chromodynamics (QCD).  A heavy quark is one whose mass, $m_Q$, is
large compared with $\Lambda_{QCD} \sim 0.3$, so that the running QCD coupling
$g_s(M_Q)$ and the associated quantity $\alpha_s(m_Q) = g_s(M_Q)^2/(4\pi)$ are
reasonably small, allowing perturbative treatments of at least some parts of
the physics of $Q \bar Q$ states and decays.  Furthermore, for
$m_Q \gg \Lambda_{QCD}$, one can obtain an approximate description of many
properties of the $Q \bar Q$ states using nonrelativistic methods, including
potential models.  From the time of the discovery of
the $J/\psi$ at BNL \cite{Aubert:1974js} and SLAC/SPEAR \cite{psi_slac} in
1974, and the $\Upsilon$ at Fermilab in 1977 \cite{Herb:1977ek,Innes:1977ae},
there has been a steadily growing wealth of data on the various $Q\bar Q$
states, where the $Q$ denotes a charm quark $c$ or a bottom/beauty quark $b$,
as well as data on mesons and baryons with charm and
bottom/beauty quantum numbers.  Some reviews of heavy quarkonia and
references to the literature include \cite{quiggrosner79}-\cite{rosner2013}.

These experimental achievements motivate the continued theoretical study of the
structure and properties of $c \bar c$ and $b \bar b$ quarkonium states.  Among
quarkonium decays, radiative decays are particularly valuable as tests of
various models, since the photon is directly observed and the nature of the
electromagnetic transition is well understood. One of the simplest types of
radiative decays is the electric dipole (E1) transition between a $Q \bar Q$
state with radial quantum number $n$ and spectroscopic type $n \, {}^{2S+1} \,
L_J =n {}^3 \, L_J$ with $L=1$ (P-wave) and $J=0, \ 1, \ 2$, denoted
$\chi_{QJ}(nP)$ in standard notation, where $Q=c, \ b$, and a lower-lying $Q
\bar Q$ S-wave state $n' \, {}^3S_1$, in particular, $J/\psi$ and
$\psi(2S)$ for the $c \bar c$ system, and $\Upsilon(n'S)$ with $n'=1, \ 2, \ 3$
for the $b \bar b$ system. In terms of $J^{PC}$ values, these decays are of the
form $J^{++} \to 1^{--} + \gamma$, where $J=0, \ 1, \ 2$.  The P-wave $c\bar c$
states were first observed in 1976 by the SLAC-LBL experiment at SLAC/SPEAR
\cite{Whitaker:1976hb,Biddick:1977sv}.  The P-wave $b \bar b$ states were first
observed by the Columbia-Stony Brook (CUSB) experiment at the Cornell CESR
$e^+e^-$ storage ring \cite{klopfenstein83,pauss83} and confirmed by the CLEO
experiment at CESR \cite{haas84}.  Larger data samples and quite accurate
measurements of branching ratios for radiative decays of P-wave $b \bar b$
states were obtained later, in particular, by the CLEO III \cite{kornicer2011}
and BABAR experiments \cite{babar2014}.  Valuable results have also been
obtained from hadron colliders, including the observation of the
$\chi_{bJ}(3P)$ $b \bar b$ states at the Large Hadron Collider (LHC)
\cite{Aad:2011ih} and the measurement of the mass of $\chi_{b1}(3P)$ by LHCb
\cite{Aaij:2014hla}.

There have been a number of theoretical studies of these
E1 transitions based on a range of different models
\cite{Karl:1980wm}-\cite{Brambilla:2005zw}.  Many of these models make use of
nonrelativistic potentials, such as the potential $V = -(4/3)\alpha_s(m_Q)/r +
\sigma r$, where the first term is a non-Abelian Coulomb potential representing
one-gluon exchange at short distances and the second term is the linear
confining potential, with $\sigma$ denoting the string tension.  These are
reasonable models, since a $Q \bar Q$ bound-state system is nonrelativistic if
$m_Q/\Lambda_{QCD} \gg 1$.

It is of interest to investigate these radiative decays of P-wave quarkonium
states using a different type of model, namely the light-front quark model
(LFQM)\cite{Terentev:1976jk}-\cite{Cheng:2003sm}.  This approach permits a
fully relativistic treatment of the quark spins and the internal motion of the
constituent quarks.  In this covariant approach, the hadronic structure for
small momentum transfer is represented by one-loop diagrams evaluated on the
light cone. It has been used to study semileptonic and nonleptonic decays of
heavy-flavor $D$ and $B$ mesons and also to evaluate radiative decay rates of
heavy mesons \cite{Hwang:2006cua,Choi:2007se,Hwang:2010iq,Ke:2010vn,Ke:2013zs}.
In particular, in \cite{Ke:2013zs} with Ke and Li, we used this approach to
calculate the widths for the radiative decays of heavy $0^{++}$ and $1^{+-}$ $Q
\bar Q$ mesons.

In the light-front formalism, one chooses the coordinate where $q^{+}=0$, in
which the quark current cannot create or annihilate pairs, and the relevant
transition matrix element can be computed as an overlap of Fock-space
wavefunctions. The terms involving pair production or annihilation vanish
\cite{Drell:1969km,Brodsky:1997de}.  An advantage of the
light-front quark model is that it is manifestly covariant. In the light-front
approach, it is easy to boost a hadron bound states from one inertial Lorentz
frame to another one when the bound state wavefunction is known in a particular
frame\cite{Brodsky:1997de}.

In this paper, extending our previous work with Ke and Li in Ref.
\cite{Ke:2013zs}, we study the radiative decays
\beq
\chi_{c1}(1P) \to J/\psi + \gamma
\label{chic1}
\eeq
and
\beq
\chi_{b1}(nP) \to \Upsilon(n'S) + \gamma
\label{chibj}
\eeq
where $n \ge n'$ by using the light-front quark model. With the front-front
formalism, we perform a numerical calculation the widths for these decays and
then compare the results with theoretical calculations based on other
approaches.

The paper is organized as follows: In Section \ref{LFQM}, we derive the
formulas for the radiative decay $1^{++} \to 1^{--} +\gamma$. Then in section
\ref{WF}, we discuss the meson wavefunctions that are relevant to the
light-front approach. In Section \ref{ANALYSIS}, we discuss numerical results
for the decay widths of $\chi_{c1}(1P) \to J/\psi + \gamma$ and
$\chi_{b1}(nP) \to \Upsilon(n'S) + \gamma$. Our conclusions are given in
Section \ref{CON}.


\section{Light-front formalism for the decays $1^{++}  \to 1^{--} +\gamma$}
\label{LFQM}

\subsection{Notation}

Here we briefly summarize the notation that is relevant for radiative
transition of meson. We follow the covariant light-front approach of
\cite{Jaus:1999zv,Cheng:2003sm} and use the same notation. In light-front
coordinates, a (four-)momentum $p$ is expressed as
\beq
p^\mu = (p^{-},p^{+},{\vec p}_{\perp})
\label{p}
\eeq
where
\beq
p^{+}=p^0+p^3, \quad p^{-}=p^0-p^3
\label{ppm}
\eeq
and
\beq
{\vec p}_{\perp}=(p^1,p^2) \ .
\label{pperp}
\eeq
Thus,
\beqs
p^2 & = & (p^0)^2 - |\vec p|^2 = (p^0)^2 - (p^3)^2 - |{\vec p}_\perp|^2 \cr\cr
    & = & p^+ p^- -|{\vec p}_\perp|^2 \ .
\label{psq}
\eeqs
We denote the momentum of the parent (incoming) meson as $P'=p'_1 +p_2$, where
$p'_{1}$ and $p_2$ are the momenta of the constituent quark and anti-quark,
with mass $m'_1$ and $m_2$, respectively.  Similarly, we label the momentum of
the daughter (outgoing) quarkonium meson as $P''=p''_1+p_2$, where
$p''_{1}$ is the momentum of the constituent quark, with mass $m''_1$.  For our
application to $Q \bar Q$ quarkonium systems, $m'_1=m_2 = m''=m_Q$.  The
four-momentum of the parent meson with mass $ M'$, in terms of light-front
coordinates, is
\beq
P'=(P'^{-},P'^{+},{\vec P}'_{\perp})
\label{pp}
\eeq
so $P'^2=P'^{+}P'^{-}-|{\vec P}_{\perp}|^2=M'^2$. Similarly, for the outgoing
meson, $P''^2=M''^2$.  In what follows, the vector signs on transverse momentum
components are to be understood implicitly and are suppressed in the notation.
The internal motion of the constituents can be described
by the variables $(x_2,p'_{\perp})$, where
\beqs
&&p'^{+}_1=x_1P'^{+}, \quad  p^{+}_2=x_2P'^{+} \nonumber\\
&&p'_{1\perp}=x_1P'_{\perp}+p'_{\perp}, \quad
  p_{2\perp}=x_2P'_{\perp} - p'_{\perp} \nonumber\\
&&x_1 + x_2 =1
\label{parts}
\eeqs
and $p''_{\perp}$ can be expressed as
\beq
p''_{\perp}=p'_{\perp}-x_2 q_{\perp} .
\label{ppp}
\eeq
%


\subsection{Form factors}

Let us define $P=P'+P ''$ and $q=P'-P''$.  Since the
initial P-wave $1^{++}$ $Q\bar Q$ state is an axial-vector, we denote it as
$A$, while the final $1^{--}$ $Q \bar Q$ state is a vector, denoted $V$.
The general amplitude for the transition $A (1^{++}) \to V(1^{--}) +\gamma$
has the form
\beq
i{\cal A} \left(A (P') \to  V(P'') \gamma (q) \right)
  =
\varepsilon^{*}_{\mu}(q) \varepsilon'_{\nu}(P')\varepsilon''^{*}_{\rho}(P'')i
\tilde {\cal A }^{\mu\nu\rho} \ ,
\label{amp}
\eeq
where $\varepsilon'_{\nu}(P')$, $\varepsilon''^{*}_{\rho}(P'')$, and
$\varepsilon^{*}_{\mu}(q)$ are the polarization (four-)vectors of the parent
heavy axial-vector meson, the daughter heavy vector meson, and the photon,
respectively.  The structure of this amplitude was given in
\cite{Dudek:2006ej}.  We review this next. Since quantum
electrodynamic (QED) interactions are invariant under parity and time reversal
(and thus also CP), this amplitude must be P- and T-invariant. In addition to
these two conditions, the transverse properties of the polarization vectors
yield the three further conditions
\beq
\varepsilon'_{\nu}(P') (P+q)^{\nu}=0,
\label{edota}
\eeq
\beq
\varepsilon''^{*}_{\rho}(P'') (P-q)^{\rho}=0,
\label{edobv}
\eeq
and
\beq
\varepsilon^{*}_{\mu}(q)q^{\mu}=0.
\label{ephoton}
\eeq
Condition (\ref{ephoton}) is also implied by electromagnetic gauge invariance.
Applying these conditions, we obtain the following general amplitude (to be
simplified below):
\beqs
i\tilde {\cal A }^{\mu\nu\rho} &=&
f_1 \epsilon^{\mu\nu\rho\alpha}P_{\alpha} +
f_2 \epsilon^{\mu\nu\rho\alpha}q_{\alpha} +
f_3 \epsilon^{\rho\nu \alpha \beta}P^\mu P_\alpha q_\beta\nonumber\\
&+&
f_4 \epsilon^{\mu\nu \alpha \beta}P^\rho P_\alpha q_\beta +
f_5 \epsilon^{\rho\mu \alpha \beta}P^\nu P_\alpha q_\beta  \ .
\eeqs
This expression can be simplified by using the fact that the photon only has
two transverse polarization states, so the timelike component
$\varepsilon^{*}_{0}(q)=0$. Taking the parent axial vector meson $A(P')$ to be
in its rest frame, we have $P'^{\mu}=(M',0)$, where $M'$ is mass of $A(P')$.
The $f_3$ term can be eliminated:
\beqs
&&[f_3 \epsilon^{\rho\nu \alpha \beta}P^\mu P_\alpha q_\beta]
\varepsilon^{*}_{\mu}(q) \varepsilon'_{\nu}(P')\varepsilon''^{*}_{\rho}(P'')
 \propto  \varepsilon^{*}_{\mu}(q)P^{\mu} \nonumber\\
&&= 2\varepsilon^{*}_{\mu}(q)P'^{\mu}=2\varepsilon^{*}_{0}(q)P'^{0}=0 \ .
\label{f3z}
\eeqs
Furthermore, the $f_1$ term vanishes due to electromagnetic gauge invariance,
$q_{\mu}\tilde A^{\mu\nu\rho}=0$. Therefore, the general
amplitude that satisfies the five conditions above is given by
\cite{Dudek:2006ej}:
\beqs
&&i\tilde {\cal A }^{\mu\nu\rho} \to i { A }^{\mu\nu\rho} \nonumber\\
&& =
f_2 \epsilon^{\mu\nu\rho\alpha}q_{\alpha} +
f_4 \epsilon^{\mu\nu \alpha \beta}P^\rho P_\alpha q_\beta +
f_5 \epsilon^{\rho\mu \alpha \beta}P^\nu P_\alpha q_\beta  \ . \nonumber \\
\label{formfactor}
\eeqs
The $f_2$ term corresponds to the electric dipole (E1) transition and makes the
dominant contribution to the amplitude, while the $f_4$ and $f_5$ terms
correspond to the magnetic dipole (M2) transition and make subdominant
contributions \cite{Cho:1994gb,Shao:2012fs}. A detailed analysis of parity and
time-reversal invariance of this general amplitude is given in
Appendix \ref{PT invariance}.

\begin{figure}
\begin{center}
\resizebox{0.45\textwidth}{!}{%
\includegraphics{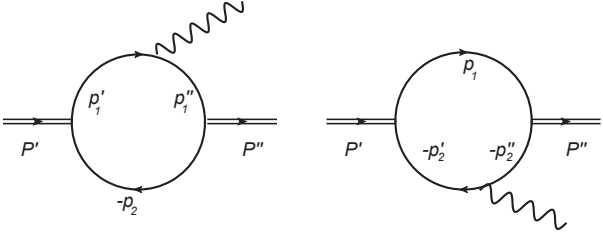}
}%
\caption {Feynman diagrams for radiative transitions in the light-front
framework.}
\label{p1}
\end{center}
\end{figure}


\subsection{Calculation of radiative decay amplitude}

In general, the width for an electromagnetic dipole transition between an
initial $Q \bar Q$
state $n \, {}^3P_J$ and a final state $n' \, {}^3S_1 + \gamma$
is given by (e.g., \cite{kwong_rosner_quigg1987})
\beq
\Gamma(n \, {}^3P_J \to n' \, {}^3S_1 + \gamma) =
\frac{4\alpha_{em} e_Q^2 E_\gamma^3}{9} \, |\langle f | {\vec r} | i \rangle
|^2 \ ,
\label{gammageneral}
\eeq
where $e_Q$ is the quark of the quark $Q$, $E_\gamma$ is the energy of the
outgoing photon in the parent rest frame, and $i$ and $j$ denote the initial-
and final-state wavefunctions.  For our calculation in the LFQM, we note that
the vertex function for the parent axial-vector meson $A(1^{++})$ is given
by \cite{Cheng:2003sm}:
\beq
iH'_{A} \left [ \gamma^{\nu} + \frac{(p'_1-p_2)^{\nu}}{W'_{A}}  \right] \gamma^5 ,
\eeq
and the vertex function for the daughter vector meson $V(1^{--})$ is
\beq
iH''_{V} \left [ \gamma^{\rho} - \frac{(p''_1-p_2)^{\rho}}{W''_{V}} \right] \ ,
\eeq
where $H'_{A}$ and $H''_{V}$ are functions of $p'_1$, $p''_1$ and $p_2$.
The explicit forms for these vertex functions will be discussed below.

There are two diagrams that contribute at leading order to the
$A \to V + \gamma$ transition amplitude, so we write
\beq
i{\cal A }^{\mu\nu\rho}(A \to V + \gamma)=i{\cal A }^{\mu\nu\rho}(a)+i{\cal A }^{\mu\nu\rho}(b) \ ,
\eeq
where the left-hand diagram in Fig.(\ref{p1}) corresponds to
${\cal A }^{\mu\nu\rho}(a)$ and the right-hand diagram in Fig.(\ref{p1})
corresponds to ${\cal A }^{\mu\nu\rho}(b)$.  These are related by charge
conjugation. For the left-hand diagram, the transition amplitude is given by
\beq
i{\cal A }^{\mu\nu\rho}(a)= e N_{e1'}  \frac{ N_c}{(2\pi)^4} \int d^4p'_1 \frac{H'_{A} H''_{V}}{N'_1N''_1 N_2} {\cal S}^{\mu\nu \rho}_a \ ,
\eeq
where
\beqs
{\cal S}^{\mu\nu \rho}_a  &=& \text {Tr} \left\{ (\pslash''_1 +m''_1)\gamma^{\mu} (\pslash_1' +m'_1)[\gamma^{\nu}+\frac{(p'_1-p_2)^{\nu}}{W'_{A}}]\right. \nonumber \\
&\times& \left. \gamma^{5} (-\pslash_2 + m_2) [\gamma^{\rho}-\frac{1}{W''_{V}}(p''_1-p_2)^{\rho}] \right\} \nonumber \\
&=& -p''^{\mu}_1 \text{Tr} [\pslash'_1 \gamma^{\nu}\pslash_2\gamma^{\rho}\gamma^5  ] + (p'_1 \cdot p''_1 - m'_1 m''_1)\nonumber\\
&\times & \text{Tr} [ \gamma^{\mu}\gamma^{\nu}\pslash_2 \gamma^{\rho}\gamma^5  ] - p'^{\mu}_1 \text{Tr}[\pslash''_1\gamma^{\nu}\pslash_2\gamma^{\rho} \gamma^5] \nonumber \\
&-& \frac{m''_1}{W'_{A}}(p'_1-p_2)^{\nu} \text {Tr} [\gamma^{\mu}\pslash'_1\pslash_2\gamma^{\rho}\gamma^5] - m''_1m_2\nonumber\\
&\times& \text {Tr} [\gamma^{\mu}\pslash'_1\gamma^{\nu}\gamma^{\rho}\gamma^5] -      \frac{m'_1}{W'_{A}}(p'_1-p_2)^{\nu} \text {Tr} [\pslash''_1\gamma^{\mu}\pslash_2\gamma^{\rho}\gamma^5] \nonumber\\
&-&m'_1m_2 \text {Tr}[\pslash''_1\gamma^{\mu}\gamma^{\nu}\gamma^{\rho}\gamma^5]-\frac{m'_1}{W''_{V}}(p''_1-p_2)^{\rho} \nonumber\\
&\times&  \text {Tr}[\pslash''_1\gamma^{\mu}\gamma^{\nu}\pslash_2\gamma^5]- \frac{m''_1}{W_{V}''}(p''_1-p_2)^{\rho} \text {Tr}[\gamma^{\mu}\pslash'_1\gamma^{\nu}\pslash_2\gamma^5] \nonumber\\
&-&[\frac{(p'_1-p_2)^{\nu}}{W'_{A}}m_2+p_2^{\nu}] \text {Tr}[\pslash''_1\gamma^{\mu}\pslash'_1 \gamma^{\rho}\gamma^5] \nonumber\\
&-&[\frac{(p'_1-p_2)^{\nu}}{W'_{A}}\frac{(p''_1-p_2)^{\rho}}{W''_{V}}-g^{\rho\nu}]  \text {Tr}[\pslash''_1\gamma^{\mu}\pslash'_1\pslash_2\gamma^5]  \nonumber\\
&-&[p^{\rho}_2+\frac{m_2}{W''_{V}}(p''_1-p_2)^{\rho}]\text
{Tr}[\pslash''_1\gamma^{\mu}\pslash'_1\gamma^{\nu}\gamma^5]  \ ,
\label{sintegral}
\eeqs
\beq
N'_1=p'^2_1-m'^2_1+i\epsilon,
\label{nprime}
\eeq
\beq
N''_1=p''^2_1-m''^2_1+i\epsilon,
\label{ndoubleprime}
\eeq
\beq
N_2=p^2_2-m^2_2+i\epsilon,
\label{n2}
\eeq
and $N_{e1'(e2)}$ denotes the electric charge of the constituent quark in units
of $e$.  The contribution to the amplitude from the right-hand diagram follows
from this.

To calculate the amplitude in the covariant light-front approach, we need to
integrate over the internal momentum, $p'^{-}_1$. In order to do this, we first
express the amplitude in terms of internal momentum $p'_1$ and external
momenta $P$ and $q$, as well as $N'_1$, $N''_1$, $N_2$, by using the
following relations:
\beqs
p''_1 &=& p'_1-q \nonumber\\
p_2 &=& (P+q)/2-p'_1 \nonumber\\
2p'_1\cdot p_2 &=& M'^2-N'_1-m'^2_1-N_2-m^2_2  \nonumber\\
2p''_1\cdot p_2 &=& M''^2-N''_1-m''^2_1-N_2-m^2_2   \nonumber\\
2p'_1\cdot p''_1 &=& -q^2 + N'_1+m'^2_1+N''_1+m''^2_1.
\label{relations}
\eeqs
After the integration over $p'^{-}_{1}$, one makes the following replacement
\cite{Jaus:1999zv,Cheng:2003sm}:
\beqs
\int d^4p'_1 \frac{H'_{A} H''_{V}}{N'_1N''_1 N_2} {\cal S}^{\mu\nu \rho}_a \varepsilon^{*}_{\mu} \varepsilon'_{\nu}\varepsilon''^{*}_{\rho}\to \nonumber \\
-i \pi \int dx_2 d^2p'_{\perp} \frac{h'_{A}h''_{V}}{x_2 \hat N'_1\hat
  N''_1}\hat{ \cal S}^{\mu\nu \rho}_a \hat \varepsilon^{*}_{\mu} \hat
\varepsilon'_{\nu} \hat \varepsilon''^{*}_{\rho} \ ,
\label{replacement}
\eeqs
where
\beqs
&&N'_1 \to \hat N'_1= x_1 (M'^2-M'^2_{0}) \nonumber \\
&&N''_1 \to \hat N''_1= x_1 (M''^2-M''^2_{0})  \nonumber \\
&& H'_{A} \to h'_{A}=(M'^2-M'^2_{0})  \sqrt{\frac{x_1 x_2}{N_c}} \frac{\tilde M'_{0}}{4M'_{0}} \phi_{np} (x_2,p'_{\perp})\nonumber\\
&& H''_{V} \to h''_{V}=(M''^2-M''^2_{0})  \sqrt{\frac{x_1 x_2}{N_c}} \frac{1}{\sqrt{2}\tilde M''_{0}} \phi_{n's} (x_2,p''_{\perp}) \nonumber\\
&&W'_{A} \to w'_{A} =  \frac{\tilde M'^2_0}{m'_1 - m_2} \nonumber\\
&& W''_{V} \to w''_{V} = M''_{0} + m''_1 + m_2 \ .
\label{replacement_relations}
\eeqs
In the above
expressions, $\phi_{n's}$ and $\phi_{np}$ represent the wavefunction for the
S-wave $Q\bar Q$ meson $V$ and the P-wave $Q\bar Q$ meson $A$,
respectively. We will discuss these wavefunctions in detail in the
next section. The definitions of $M'_0$, $M''_0$, $\tilde M'_0$ and $\tilde
M''_0$ are given in appendix \ref{experssion}. The definition of $\hat
\varepsilon^{*}$, $\hat \varepsilon'$ and $\hat \varepsilon''^{*}_{\rho}$ is
given in \cite{Jaus:1999zv,Cheng:2003sm}.

One should also include the contribution from zero modes in the $A$ meson.
In practice, this amounts to the following replacement for
$p'_{1\mu}$ in $\hat{ \cal  S}^{\mu\nu \rho}_a $ in the integral
\cite{Jaus:1999zv,Cheng:2003sm}:
\beqs
\hat p'_{1\mu} &\to& P_{\mu} A^{(1)}_1 + q_\mu A^{(1)}_2 \ , \nonumber\\
 \hat p'_{1\mu} \hat p'_{1\nu}  &\to &g_{\mu\nu} A^{(2)}_1 + P_{\mu}P_{\nu} A^{(2)}_2  \nonumber\\
&+& (P_{\mu}q_{\nu} + q_{\mu}P_{\nu})A^{(2)}_3 + q_{\mu} q_{\nu} A^{(2)}_4 \ , \nonumber \\
 \hat p'_{1\mu} \hat p'_{1\nu}  \hat p'_{1\alpha}  &\to& (g_{\mu\nu}P_{\alpha} + g_{\mu\alpha} P_{\nu} + g_{\nu\alpha}P_{\mu})A^{(3)}_1 \nonumber\\
 &+& (g_{\mu\nu}q_{\alpha} + g_{\mu\alpha} q_{\nu} + g_{\nu\alpha}q_{\mu})A^{(3)}_2 \nonumber\\
 &+& P_{\mu}P_{\nu} P_{\alpha} A^{(3)}_3 + (P_{\mu}P_{\nu}q_{\alpha} + P_{\mu}q_{\nu}P_{\alpha} \nonumber\\
 &+& q_{\mu} P_{\nu} P_{\alpha}) A^{(3)}_4 + (q_{\mu}q_{\nu}P_{\alpha} + q_{\mu}P_{\nu}q_{\alpha}  \nonumber\\
 &+& P_{\mu}q_{\nu} q_{\alpha}) A^{(3)}_5 + q_{\mu} q_{\nu} q_{\alpha} A^{(3)}_6 \ .
\eeqs
After these operations, the amplitude ${\cal A }^{\mu\nu\rho}(a)$ can be
expressed as a function of the external four-momenta $P$ and $q$. It can be
parametrized in the following form:
\beqs
i{\cal A }^{\mu\nu\rho}(a) &=&   f^a_2 \epsilon^{\mu\nu\rho\alpha}q_{\alpha} +  f^a_4 \epsilon^{\mu\nu \alpha \beta}P^\rho P_\alpha q_\beta \nonumber\\
&+& f^a_5 \epsilon^{\rho\mu \alpha \beta}P^\nu P_\alpha q_\beta \  ,
\eeqs


with
\begin{widetext}
\beqs
f^a_2  (q^2) &=&  e N_{e'_1} \frac{N_c}{16\pi^3} \int dx_2 d^2p'_{\perp} \frac{h'_{A}h''_{V}}{x_2\hat N'_1 \hat N''_1} (-4)\cdot\left [  \frac{1}{w'_{A}}(m''_1+ m'_1-2m_2) A^{(2)}_1 + \frac{1}{w''_{V}}(2m_2+m'_1+ m''_1) A^{(2)}_1   \right. \nonumber\\
&-& \left. \frac{1}{4}(1-2A^{(1)}_2)\left (-q^2+\hat N'_1  + \hat N''_1 + (m'_1 - m''_1)^2\right) - A^{(1)}_{2} (m''_1 m_2 - m'_1 m_2) -  m'_1 m_2 \right ] \ , \nonumber\\
f^a_4  (q^2)&=&   e N_{e'_1} \frac{N_c}{16\pi^3} \int dx_2 d^2p'_{\perp} \frac{h'_{A}h''_{V}}{x_2\hat N'_1 \hat N''_1} (-4)\cdot\left [
\frac{1}{w''_{V}}\left ( (m'_1 -m''_1) (A^{(2)}_3 + A^{(2)}_4 - A^{(1)}_2) + (m'_1 +m''_1 + 2m_2)  \right.  \right.\nonumber\\
&\times& \left.\left. (A^{(2)}_2 + A^{(2)}_3 -A^{(1)}_1) - m'_1 (A^{(1)}_1 + A^{(1)}_2 -1) \right) - A^{(1)}_1 + A^{(2)}_2 + A^{(2)}_3 - \frac{1}{w'_{A}w''_{V}}(2A^{(3)}_1 + 2A^{(3)}_2 - 2A^{(2)}_1) \right] \nonumber\\
f^a_5  (q^2)&=&   e N_{e'_1} \frac{N_c}{16\pi^3} \int dx_2 d^2p'_{\perp} \frac{h'_{A}h''_{V}}{x_2\hat N'_1 \hat N''_1} (-4)\cdot\left [
\frac{1}{w'_{A}}\left ( (m'_1 -m''_1) (A^{(2)}_3 - A^{(2)}_4) + (m'_1 + m''_1 - 2m_2)  \right.  \right.\nonumber\\
&\times& \left.\left. (A^{(2)}_2 - A^{(2)}_3 ) + m'_1 (A^{(1)}_2 - A^{(1)}_1 ) \right)  + A^{(2)}_2- A^{(2)}_3 - \frac{1}{w'_{A}w''_{V}}( 2A^{(3)}_2 -2A^{(3)}_1) \right]  \ ,
\label{ffexpression}
\eeqs

\end{widetext}
where the explicit expression of $A^{(i)}_j$ is given in Appendix
\ref{experssion}.

For the right-hand diagram in Fig.(\ref{p1}), the amplitude ${\cal A
}^{\mu\nu\rho}(b)$ can be obtained from ${\cal A }^{\mu\nu\rho}(a)$ by the
interchanges $m'_1 \leftrightarrow m'_2$, $m''_1 \leftrightarrow m''_2$, $m_2
\leftrightarrow m_1$, $N_{e'_1} \leftrightarrow N_{e_2} $:
\beqs  i {\cal A
}^{\mu\nu\rho}(b) &=& f^b_2 \epsilon^{\mu\nu\rho\alpha}q_{\alpha} +
f^b_4 \epsilon^{\mu\nu \alpha \beta}P^\rho P_\alpha q_\beta  \nonumber\\
&+& f^b_5 \epsilon^{\rho\mu \alpha \beta}P^\nu P_\alpha q_\beta \ .
\label{interchange_amplitude}
\eeqs
The coefficients $f_i$ in Eq.(\ref{formfactor}) are the sum of contributions
from two parts, ${\cal A }^{\mu\nu\rho}(a)$ and ${\cal A }^{\mu\nu\rho}(b)$:
\beq f_i(q^2) =f^a_i (q^2) + f^b_i (q^2) , \quad (i=2, 4, 5) \ .
\label{interchange_amp}
\eeq
In Eqs. (\ref{ffexpression}) and (\ref{interchange_amp}) we write these as
general form factors dependent on $q^2$, but note that in the physical $A \to V
\gamma$ decay, $q^2=0$ for the real outgoing photon, so that these are simply
constant coefficients.  We use this generalization to nonzero $q^2$ because in
the light-front formalism, these form factors are calculated in the region
where the photon momentum is not onshell, i.e., where $q^2 \ne 0$.
To obtain the physical
values $f_i(0)$ and calculate the decay rate, we take limit $q^2 \to 0$.  This
yields the resulting width
\begin{widetext}
\beqs
\Gamma(n{}^3P_1 \to n'{}^3S_1) &=& \frac{q^3}{24\pi}\left \{ \frac{2}{M''^2}[f^2_2 +4f_2f_4 M' q + 4 M'^2 q^2 f^2_4]  + \frac{2}{M'^2}[ f^2_2 -4f_2 f_5  M' q+ 4 M'^2 q^2 f^2_5 ] \right\}   \ ,
\label{widthLF}
\eeqs
\end{widetext}
where $q= (M'^2-M''^2)/(2M')$ is the momentum of the emitted photon.  In this
paper we focus on E1 dipole transition rates, which are dominant, and hence
drop the subdominant $f_4$ and $f_5$ terms in the calculations.


\section{Wavefunctions for heavy quarkonium states}
\label{WF}

The wavefunctions $\phi_{n's}$ and $\phi_{np}$ can, in principle, be derived
from relativistic light-front Bethe-Salpter type equations
\cite{Jaus:1989au,Cheung:1995ub}. However, as discussed in Refs.
\cite{Jaus:1989au} and \cite{Isgur:1988gb}, there is a simpler approach, namely
to use wavefunctions from nonrelativistic quark models with given potentials.
Although a QCD-motivated potential has the form $V = -(4/3)\alpha_s(m_Q)/r +
\sigma r$, as noted above, this involves the complication of requiring
numerical solutions of the Sch\"odinger equation.  To avoid this complication,
Refs. \cite{Jaus:1989au} and \cite{Isgur:1988gb} used variational solutions of
the Schr\"odinger equation with a nonrelativistic harmonic oscillator
potential. This approach was also adopted by
Refs. \cite{Jaus:1999zv,Cheng:2003sm,Choi:1997iq,Choi:1999nu}.  However, the
predictions from this type of approach do not fit the measured widths for
$\Upsilon (nS)$ well, and to overcome this problem, modified harmonic
oscillator wavefunctions were suggested in \cite{Ke:2010vn}. The normalization
and explicit expressions for the modified harmonic wavefunctions are listed in
Appendix \ref{wavefunction}.

In the next section, we use the modified harmonic oscillator wavefunctions in
\cite{Ke:2010vn} to calculate numerically the radiative decay widths of
$\chi_{c1}(1P)$ and $\chi_{b1}(nP)$ states and to compare these with
theoretical predictions from other models.

Some comments are appropriate concerning approaches other than the light-front
approach.  For the heavy quarkonium system, nonrelativistic potential models
such as Cornell potential model have proved to be generally rather successful
in fitting data
\cite{Eichten:1978tg,Eichten:1979ms,Eichten:1976jk,Eichten:1994gt,Buchmuller:1980su}. There
are also analyses of relativistic corrections to potential models, such as
\cite{Gupta:1982kp,Moxhay:1983vu,Kwong:1988ae}. A relativistic quark model was
proposed in Ref. \cite{Godfrey:1985xj}. Screening effects were studied in
Refs. \cite{Laermann:1986pu,Chao:1992et,Ding:1993uy}, and additional potential
models were used in \cite{Sumino:2001eh,Recksiegel:2001xq}. In these potential
models, the wavefunctions can be obtained by numerically solving the
Schr\"odinger equations.  In future work it would be of interest to
investigate the differences in radiative widths calculated using the
phenomenological wavefunctions for the light-front quark model adopted
here (with modified harmonic oscillator wavefunctions) and wavefunctions from
potential models. Here we focus on calculations using modified harmonic oscillator
wavefunctions, and we compare these with results obtained from other
approaches.


\section{Analysis of radiative transitions of $\chi_{c1}(1P) $ and
$\chi_{b1}(nP)$ }
\label{ANALYSIS}

\begin{widetext}

\begin{table}
\begin{ruledtabular}
\caption{Decay width (in units of keV) of
$\chi_{c1}(1P) \to J/\psi + \gamma $ in the light-front approach, based on
modified harmonic oscillator wavefunctions \cite{Ke:2010vn}.
The predictions from other models
(relativistic quark model\cite{Ebert:2002pp,Godfrey:2015dia},
nonrelativistic screened potential model \cite{Brambilla:2004wf})
and experimental data from PDG \cite{PDG} are also listed for comparison.
The parameters sets are CM1 and CM2. We use the PDG fitted value
$\Gamma_{\chi_{c1}}=840 \pm 40$ keV and $BR(\chi_{c1}(1P) \to J/\psi+\gamma) =
33.9 \pm 1.2$ \% \cite{PDG}. For the entry referring to Ref.
\cite{Ebert:2002pp}, we list three values presented there, based on the
specific models used in that work.} \label{tabCM}
\begin{tabular}{ccccccc}
  Decay mode & CM1 & CM2 & exp.(PDG)\cite{PDG}  & \cite{Brambilla:2004wf}(NR)  & \cite{Ebert:2002pp}  \\
  \hline
$\chi_{c1}(1P) \to J/\psi+\gamma$ & $324 \pm 20$ & $282 \pm 35$&
$285 \pm 30$ & 241  & 265/285/305  \\
\end{tabular}
\end{ruledtabular}
\end{table}

\begin{table}
\begin{ruledtabular}
\caption{Coefficients $f_2$, $f_4$, and $f_5$ for
$\chi_{b1}(nP) \to \Upsilon (n'S) \gamma$ in covariant light-front approach,
where $f_i \equiv f_i(q^2=0)$.}    \label{tabFF}
\begin{tabular}{ccccccc}
  Decay mode &  $f_2$  & $f_4 (\text{GeV}^{-2})$ & $f_5 (\text{GeV}^{-2})$  \\
  \hline
   $\chi_{b1}(1P) \to \Upsilon (1S) \gamma$ &  $-0.94 \pm0.06 $ & $0.0049 \pm 0.0004$ &$-0.0083\pm0.0002$ \\
   $\chi_{b1}(2P) \to \Upsilon (1S) \gamma$ & $ +0.21 \pm0.05$ &  $0.0019\pm0.0006$ &$0.0037  \pm0.0002$ \\
   $\chi_{b1}(2P) \to \Upsilon (2S) \gamma$ &  $- 1.26\pm0.10$ &  $0.0094\pm0.0010$ &$-0.0071 \pm0.0008$ \\
   $\chi_{b1}(3P) \to \Upsilon (1S) \gamma$ & $-0.11\pm 0.03$  &  $-0.0014\pm0.0002$ &$-0.0021\pm0.0003$ \\
   $\chi_{b1}(3P) \to \Upsilon (2S) \gamma$ & $+0.29 \pm 0.10 $ & $0.0038\pm0.0016$ &$0.0050  \pm0.0002$ \\
   $\chi_{b1}(3P) \to \Upsilon (3S) \gamma$ &  $-1.39\pm 0.06 $ &$0.0056\pm0.0032$ &$-0.0087  \pm0.0015$ \\
\end{tabular}
\end{ruledtabular}
\end{table}

\begin{table}
\begin{ruledtabular}
\caption{Decay widths (in units of keV) of
$\chi_{b1}(nP) \to \Upsilon(n'S) + \gamma$  E1 decays in the light-front
approach, denoted $\Gamma_{\text{MSHO}}$,
based on modified simple
harmonic oscillator (MSHO) wavefunctions \cite{Ke:2010vn}. The predictions from
other models (relativistic quark model\cite{Ebert:2002pp,Godfrey:2015dia},
non-relativistic screened potential model \cite{Li:2009nr}, and
nonrelativistic constituent quark model \cite{Segovia:2016xqb}) are also
listed for comparison, where  \cite{Li:2009nr}$_0$ denotes results from
the $SNR_0$ (screened nonrelativistic) model and \cite{Li:2009nr}$_1$ denotes results from the
$SNR_1$ model. We also list the ratio $\Gamma_{\text{MSHO}}/\Gamma_{\text {th(ave.)}}$, where $\Gamma_{\text {th(ave.)}}$ is average value of widths from other theoretical models.  }  \label{tabM}
\begin{tabular}{ccccccccc}
  Decay mode & $\Gamma_{\text{MSHO}}$   & \cite{Ebert:2002pp}  & \cite{Li:2009nr}$_0$  & \cite{Li:2009nr}$_1$ & \cite{Godfrey:2015dia} & \cite{Segovia:2016xqb}& $\Gamma_{\text{MSHO}}/\Gamma_{\text {th(ave.)}}$ \\
  \hline
   $\chi_{b1}(1P) \to \Upsilon (1S) \gamma$ &  $37.3 \pm 4.8 $ &  36.6 &33.6 &30.0 &29.5 &35.66& $1.12\pm 0.15$\\
   $\chi_{b1}(2P) \to \Upsilon (1S) \gamma$ &  $10.6 \pm 5.5 $   &7.49 &12.4&8.56 &5.5 &9.13& $1.23\pm 0.64$\\
   $\chi_{b1}(2P) \to \Upsilon (2S) \gamma$ &  $10.0 \pm 1.7 $   &14.7 &15.9&13.8&13.3 &15.89& $0.68\pm 0.12$\\
   $\chi_{b1}(3P) \to \Upsilon (1S) \gamma$ &  $6.1 \pm 3.9 $  & &6.80&3.39 &1.3 &4.17 & $1.56\pm 1.00$\\
   $\chi_{b1}(3P) \to \Upsilon (2S) \gamma$ &  $4.7 \pm 3.2$   & & 5.48&5.39 &3.1 &4.58& $1.01\pm 0.69$ \\
   $\chi_{b1}(3P) \to \Upsilon (3S) \gamma$ &  $3.6 \pm 0.4$   &  &12.0&9.97 &8.4&9.62& $0.36\pm 0.04$ \\
\end{tabular}
\end{ruledtabular}
\end{table}

\end{widetext}

In this section we apply the light-front formalism for the decay $A (1^{++} )
\to V (1^{--}) + \gamma $ to the analysis of the radiative decays
$\chi_{c1}(1P) \to J/\psi + \gamma$ and $\chi_{b1}(nP) \to \Upsilon
(n'S)+\gamma$. We present the results of our numerical calculations of form
factors (evaluated at $q^2=0$) and decay widths. For the charmonium
$\chi_{c1}(1P)$ decay, we compare our result with experimental data on the
width, as listed in the Particle Data Group Review of Particle Properties (RPP)
\cite{PDG}. Although the RPP lists this width for the decay $\chi_{c1}(1P) \to
J/\psi$, it does not list widths for the $\chi_{b1}(nP) \to \Upsilon (n'S) +
\gamma$ decays, only branching ratios.  Since our calculation yields the width
itself, and a calculation of the branching ratio requires division by the total
width in each case, we therefore compare our results on the branching ratios
for these decays with predictions from other models, including the relativistic
quark model \cite{Ebert:2002pp,Godfrey:2015dia}, the non-relativistic screened
potential model \cite{Li:2009nr}, and the nonrelativistic constituent quark
model \cite{Segovia:2016xqb}. For each decay, we have performed numerical
calculations based on modified harmonic oscillator wavefunctions as discussed
in \cite{Ke:2010vn}.

First, we study the charmonium radiative decay $\chi_{c1}(1P) \to J/\psi
+\gamma$.  The parameter sets that we use are as follows, with labels
indicated:

\begin{enumerate}

\item CM1: $m_c$ = 1.4 \ GeV, \\
 $\beta_{\chi_{C1}(J/\psi)}$ = 0.639$\pm 0.020$ \ GeV.

\item CM2:  $m_c$ = 1.5 \ GeV, \\
  $\beta_{\chi_{C1}(J/\psi)}$ = 0.600$\pm 0.020$ \ GeV.

\end{enumerate}

We present our results in Table \ref{tabCM}, with the uncertainties arising
from the uncertainties in the $\beta$ parameters, as in \cite{Ke:2013zs}.
As one can see from Table \ref{tabCM}, our results agree with experimental
data within the range of experimental and theoretical uncertainties. The
theoretical uncertainties arise from the value of $m_c$ taken and also from the
model used. The model-dependent uncertainties will be evident from our
comparison of predictions from various models.

Next, we proceed to analyze the radiative decays of P-wave $b \bar b$
states. We use the modified harmonic oscillator wavefunctions, which have been
successfully applied to the study of radiative decays of $\Upsilon (nS) \to
\eta_b + \gamma$ \cite{Ke:2010vn}.  In this case, the LFQM has the following
parameters: the mass of the quark, $m_b$, the harmonic oscillator wavefunction
parameter for $\chi_{b1}(nP)$, $\beta_{\chi_{b1(nP)}}$, and the wavefunction
parameter for $\Upsilon(nS)$ $\beta_{\Upsilon(nS)}$.  For the mass of the
quark, we use $m_b = 4.8$ GeV. This is an effective $b$-quark mass chosen to
optimize the fit to these radiative transitions, as has been done in a number
of other studies; for example, the recent comprehensive study
\cite{Godfrey:2015dia} uses the value $m_b=4.977$ GeV.

For the effective harmonic oscillator wavefunction parameters, there are two
choices. One is to use a single parameter $\beta$ for all states in the $b \bar
b$ system. In this case, the wavefunctions correspond to eigenstates of the
harmonic oscillator Hamilton with $V \propto r^2$, and hence the energy
splitting between different energy levels is \cite{quiggrosner79}
$\Delta E \propto \mu^{-1/2}$, where $\mu=m_Q/2$ is the reduced mass of $Q\bar
Q$ system. This does not account for the observed approximate equality of mass
splittings $m(\psi(2S))-m(J/\psi) \simeq m(\Upsilon(2S))-m(\Upsilon(1S))$.
Therefore, a more practical choice
is to treat $\beta$ as variational parameter to fit each state separately. For
example, in Ref. \cite{Godfrey:2015dia}, the authors obtain $\beta$ by equating
the rms radius of the harmonic oscillator wavefunction for the specified states
with the rms radius of the wavefunctions calculated using the relativized quark
model. Explicitly, for $n=1$, $\beta \sim 0.9 -1.2$ GeV, for $n=2$,
$\beta \sim 0.7 -0.8$ GeV, and for $n=3$, $\beta \sim 0.6 -0.7$ GeV.
For our modified harmonic wavefunctions, these results are not exact,
but can serve as an estimate of the range of wavefunction parameters. In our
analysis, we use the following values of wavefunction parameters:

\begin{enumerate}

\item
$\beta_{\chi_{b1}(1P)}$ = 1.00$\pm0.02$ GeV,

\item
$\beta_{\chi_{b1}(2P)}$ =  0.71$\pm0.02$ GeV,

\item
$\beta_{\chi_{b1}(3P)}$ = 0.70$\pm0.02$ GeV,

\item
$\beta_{\Upsilon(1S)}$ = 0.90$\pm0.02$ GeV,

\item
$\beta_{\Upsilon(2S)}$ = 0.71$\pm0.02$ GeV,

\item

$\beta_{\Upsilon(3S)}$ = 0.70$\pm0.02$ GeV.

\end{enumerate}

Here we have used estimated values of the uncertainties in these
parameters corresponding to those that we used in our previous study
\cite{Ke:2013zs}.  The uncertainties that we include with our resultant
calculations of radiative decay widths incorporate these uncertainties.

For the values of form factors, we show typical results in Table {\ref{tabFF}.
  The numerical results for the decay widths calculated with our parameter
  setting are listed in Table {\ref{tabM}.  In both of these tables, we include
    the estimated uncertainties arising from the uncertainties in the input
    value of $m_b$ and the input values of the $\beta$ parameters. Since for
    $\chi_{b1}(nP)$ system, only branching ratios are experimentally
    determined, we compare our results, denoted $\Gamma_{\text {MSHO}}$, with
    those from other theoretical models. As an rough estimation, we define the
    average values of widths from these theoretical models
    \cite{Ebert:2002pp,Godfrey:2015dia,Li:2009nr,Segovia:2016xqb} to be
    $\Gamma_{\text{th(ave.)}}$.  It should be noted that for many of the decay
    modes, there is a substantial spread of values of branching ratios
    predicted by different models. We then calculate the ratio $\Gamma_{\text
      {MSHO}}/\Gamma_{\text{th(ave.)}}$ and list this ratio in Table
    {\ref{tabFF}. The decay $\chi_{b1}(1P) \to \Upsilon(1S) + \gamma$ has a
      measured branching ratio $BR(\chi_{b1}(1P) \to \Upsilon(1S) + \gamma) =
      33.9 \pm 2.2$ \% \cite{PDG}.  For this decay mode, our predicted width
      agrees well with the average of the other models and, furthermore, the
      predictions of these other models agree well among themselves.  The
      measured branching ratios for the radiative decays of the $\chi_{b1}(2P)$
      are $BR(\chi_{b1}(2P) \to \Upsilon(2S) + \gamma) = 19.9 \pm 1.9$ \% and
      $BR(\chi_{b1}(2P) \to \Upsilon(1S) + \gamma) = 9.2 \pm 0.8$ \%
      \cite{PDG}. Our predicted width for the first of these decays is in good
      agreement with the average of the predictions of other models, while our
      predicted width for the second of these decays is slightly smaller than
      this average. $\chi_{bJ}(3P)$ states have recently been observed at the
      LHC via their radiative decays \cite{Aad:2011ih,Aaij:2014hla} (although
      no branching ratios for these decays are listed yet by the PDG). For
      radiative decays of $\chi_{b1}(3P)$ to $\Upsilon(1S)$ and $\Upsilon(2S)$,
      our LFQM predictions are in good agreement, to within uncertainties, with
      other models, while our prediction for the decay to $\Upsilon(3S)$ is
      somewhat smaller than the predictions from other models.

      In general, these results show that the light-front quark model with
      phenomenological meson wavefunctions (specifically, modified harmonic
      oscillator wavefunctions), is suitable for the calculation of $nP \to
      n'S$ radiative decay widths, since this model gives reasonable
      predictions for these widths, as compared with experimental data and
      other theoretical approaches.  The results from the calculations in the
      covariant light-front approach and corresponding
      nonrelativistic/relativized quark model calculations reflect some
      differences in the predictions of decay widths, which are related to
      differences in the properties of these respective models.  Specifically,
      nonrelativistic/relativized quark models contain different ways of
      including relativistic corrections and also truncations of these
      relativistic effects, while in the LFQM these relativistic corrections
      are systematically included.  This shows one advantage of the covariant
      light-front approach, namely, that it is a fully relativistic formalism,
      and one does not need to carry out a reduction from relativistic
      interaction terms to the nonrelativistic limit.

One drawback in the current LFQM is that we do not know the exact form of the
light-front wavefunctions and hence only use trial wavefunctions. This
problem is more serious for excited states, because for excited $b \bar b$
states with radial quantum numbers $n \ge 2$, where $\Lambda_{QCD}$ is larger
than the typical binding energy of the state, the Coulombic type potential is
no longer a very good approximation \cite{Brambilla:2010cs}, so we have larger
uncertainties in the $b \bar b$ wavefunction that serves as input in
light-front quark model.  This can be seen from Table \ref{tabM}; for reasonable parameters,
the decay width of $\chi_{b1}(1P)$ from the LFQM agrees well with predictions
 from nonrelativistic/relativized quark
models, but for excited states, the LFQM calculations for two channels do
not match perfectly with predictions from these nonrelativistic/relativized
quark models.  As been pointed out in Ref. \cite{Brambilla:2010cs}, for
radiative transition of these excited $b \bar b$ states, we rely on
phenomenological models, but these do not always agree with QCD in the
perturbative regime. Even though the LFQM is a fully relativistic approach,
there is thus motivation for further theoretical work to gain a better
understanding of the determination of light-front wavefunctions for
$Q \bar Q$ states.


\section{Conclusion}
\label{CON}

In this paper we have derived formulas for the radiative decay of $1^{++}$
heavy mesons via the channel $1^{++} \to 1^{--} +\gamma$ in the light-front
quark model. Then we have applied these to calculate the coefficients $f_i$ and
the radiative decay widths of $\chi_{c1}(1P)$ and $\chi_{b1}(nP)$ via the
respective channels $\chi_{c1}(1P) \to J/\psi + \gamma$ and
$\chi_{b1}(nP) \to \Upsilon(n'S) + \gamma$. Within the LFQM framework, we have
adopted modified harmonic-oscillator wavefunctions.
We have shown that most of the predictions of the LFQM with modified
harmonic-oscillator wavefunctions are in reasonable agreement with data and
other model calculations.


\begin{acknowledgments}
  We are grateful to Prof. Robert Shrock for his helpful suggestions and
  assistance. This research was partially supported by the NSF grant
  NSF-PHY-13-16617. We are also grateful to Profs. Hong-Wei Ke and Xue-Qian Li
  for collaboration on our previous related work \cite{Ke:2013zs}.
\end{acknowledgments}


\begin{appendix}

\section{ Time reversal and Parity transformations of  amplitude}
\label{PT invariance}

\subsection{Time Reversal Transformation}

The action of time reversal on on S-matrix element $\langle \beta | H | \alpha
\rangle $ is defined to be $(\langle \tilde\beta | {\cal T}H {\cal T}^{-1}|
\tilde \alpha \rangle)^{*}$, where $|\tilde \alpha \rangle={\cal T} | \alpha
\rangle $. So the time-reversal invariance of electromagnetic interaction is
\cite{Dudek:2006ej}:
\beq
\langle \beta | H_{\text em} | \alpha \rangle =(\langle \tilde\beta | {\cal T}H_{\text em} {\cal T}^{-1}| \tilde \alpha \rangle)^{*}=(\langle \tilde\beta | H_{\text em} | \tilde \alpha \rangle)^{*}
\label{Ttrans}
\eeq
where we use the time-reversal invariance of the electromagnetic Hamiltonian
operator: $H_{\text em}={\cal T}H_{\text em} {\cal T}^{-1}$.

For a state with 3-momentum $\vec{p}$, spin $J$ and $z$-component of spin $m$,
the time-reversal transformation is ${\cal T} |\vec{p},J,m\rangle=
\zeta(-1)^{J-m}|-\vec{p},J,-m\rangle $ (for vector and axial-vector states,
$\zeta =+1$). After contractions with the associated field operator, this
amounts to the change of polarization: $\varepsilon^{\mu}(\vec{p},m) \to
\zeta(-1)^{J-m}\varepsilon^{\mu}(-\vec{p},-m) =\zeta(-1)^{J+1} {\text
  {P}}^{\mu}_{\nu}\varepsilon^{\nu*}(\vec{p},m) $, where we have used the
relation $\varepsilon^{\mu*}(\vec{p},m)=(-1)^{m+1}{\text
  {P}}^{\mu}_{\nu}\varepsilon^{\nu}(-\vec{p},-m)$, and ${\text {P}}^{\mu}_{\nu}
= \text {diag} (1,-1,-1,-1)$ represents spatial inversion \cite{Dudek:2006ej}.

The general amplitude in Eq.(\ref{formfactor}) should satisfy the time-reversal
invariance condition in Eq.\ref{Ttrans}. Let us consider the $f_2$ term
first. Without loss of generality, we can choose polarization (+,+,+) states;
then this is given by
\beq
{\cal M}_{+++}=-i\varepsilon^{*}_{\mu}(q,+)
\varepsilon'_{\nu}(P',+)\varepsilon''^{*}_{\rho}(P'',+)f_2
\epsilon^{\mu\nu\rho\alpha}q_{\alpha} \ .
\label{f2M+++}
\eeq
In this case where the three polarization vectors are all transversal and only
carry spatial components of Lorentz indices, the index of the photon momentum
$q_{\alpha}$ has to be $\alpha=0$:
\beq
{\cal M}_{+++}=-i\varepsilon^{*}_{i}(q,+)
\varepsilon'_{j}(P',+)\varepsilon''^{*}_{k}(P'',+)f_2 \epsilon^{ijk}q_{0} \ .
\label{f2M+++2}
\eeq
Under a time-reversal transformation, $q^{0} \to q^{0}$, $\epsilon^{i}
(\epsilon'^{i}, \epsilon''^{i})$ $\to$ $(-1)\epsilon^{i*}(\epsilon'^{i*},
\epsilon''^{i*} )$ and
\beqs
{\widetilde {\cal M}}_{+++}&=& -(-1)^3\left (i\varepsilon_{i}(q,+) \varepsilon'^{*}_{j}(P',+)\varepsilon''_{k}(P'',+)f_2 \epsilon^{ijk}q_{0} \right)^{*} \nonumber\\
&=&  -i\varepsilon^{*}_{i}(q,+) \varepsilon'_{j}(P',+)\varepsilon''^{*}_{k}(P'',+)f^{*}_2 \epsilon^{ijk}q_{0}  \ .
\eeqs
According to Eq. (\ref{Ttrans}), the amplitude is time-reversal invariant if
\beq
{\cal M}_{+++} ={\widetilde {\cal M}}_{+++} \ \to \  f_2 = f^*_2 \ ,
\eeq
which is satisfied as we can see from the explicit expression of $f_2$ in
Eq.(\ref{ffexpression}). Using an equivalent analysis, we can prove that
the $f_4$ and $f_5$ terms also preserve time-reversal invariance.

\subsection{Parity Transformation}

For a physical state $| \alpha \rangle$, the action of a parity transformation
is ${\cal P} | \alpha \rangle =| \alpha' \rangle$.  The parity
invariance of the electromagnetic interaction is expressed as
\beq
\langle \beta| H_{\text {em}} | \alpha \rangle = \langle \beta|{\cal P}^{-1} {\cal P} H_{\text {em}} {\cal P}^{-1}{\cal P} | \alpha \rangle= \langle \beta'| H_{\text {em}} | \alpha' \rangle
\label{Ptrans}
\eeq
The parity transformation of a state with 3-momentum $\vec{p}$, spin $J$, and
$z$-component of spin $m$ is defined as ${\cal P}|\vec{p},J,m\rangle=
\eta_{P}|-\vec{p},J,m\rangle $, where $\eta_{P}$ is the intrinsic parity of
this state. For a vector meson, $\eta_{P}=-1$, and for an axial vector meson,
$\eta_{P}=+1$. After contractions with the associated field operator, this
amounts to the change of polarization: $\varepsilon^{\mu}(\vec{p},m) \to
\eta_{P}\varepsilon^{\mu}(-\vec{p},m) =-\eta_{P}{\text
  {P}}^{\mu}_{\nu}\varepsilon^{\nu}(\vec{p},m) $, where we have used the
relation $\varepsilon^{\mu}(-\vec{p},m) = -{\text
  {P}}^{\mu}_{\nu}\varepsilon^{\nu}(\vec{p},m) $.

The general amplitude in Eq.(\ref{formfactor}) should satisfy the parity
invariance condition in Eq.\ref{Ptrans}. We take the $f_2$ term as an example
to demonstrate this requirement. Without loss of generality, we can choose
the polarization (+,+,+) states, for which the amplitudes are given by
Eq.(\ref{f2M+++}) and Eq.(\ref{f2M+++2}).

Under a parity transformation, $q^{0}\to q^{0}$, $\epsilon^{i} (\epsilon'^{i},
\epsilon''^{i})$ $\to$ $\eta_{P}(+1)\epsilon^{i}(\epsilon'^{i}, \epsilon''^{i}
)$, the amplitude ${\cal M}_{+++}$ is transformed to
\beqs
{\cal M'}_{+++}&=&-\eta_{V}\eta_{A}\eta_{\gamma}i\varepsilon^{*}_{i}(q,+) \varepsilon'_{j}(P',+)\varepsilon''^{*}_{k}(P'',+)f_2 \epsilon^{ijk}q_{0} \nonumber\\
&=&-(+1)i\varepsilon^{*}_{i}(q,+) \varepsilon'_{j}(P',+)\varepsilon''^{*}_{k}(P'',+)f_2 \epsilon^{ijk}q_{0} \ , \nonumber\\
\label{M+++P}
\eeqs
where the intrinsic parities of $V$, $A$ and $\gamma$ are
$\eta_{V}=-1$, $\eta_{A}=+1$ and $\eta_{\gamma}=-1$, respectively. From
Eq.(\ref{M+++P}), we can see ${\cal M'}_{+++}={\cal M}_{+++}$; hence parity is
conserved for the $f_2$ term. Applying the same method of analysis, we can
prove that the $f_4$ and $f_5$ terms also preserve parity invariance.


\section{The wavefunctions}
\label{wavefunction}

The normalization of the S-wave meson wavefunction in the light-front framework
is
\beq
\frac{1}{2(2\pi)^3} \int dx_2 dp^2_{\perp} |\phi_{n's}(x_2, p_{\perp})|^2 =1.
\eeq
Here $\phi_{n's}(x_2, p_{\perp})$ is related to the wavefunction in normal
coordinates $\psi_{n's}(p)$ by
\beq
\phi_{n's} (x_2, p_{\perp})= 4\pi^{\frac{3}{2}} \sqrt{\frac{d p_z}{dx_2}}\psi_{n's} (p) \ , \quad \frac{dp_z}{dx_2}= \frac{e'_1e_2}{x_1x_2 M'_{0}} \
\eeq
The normalization of $\psi_{n's}(p)$ is given by
\beq
\int d{\bf  p}^3 |\psi_{n's}(p)|^2 = 4\pi  \int  p^2 dp |\psi_{n's}(p)|^2 =1 \ .
\eeq
The normalization for the P-wave meson wavefunction in the light-front
framework is \cite{Cheng:2003sm}
\beq
\frac{1}{2(2\pi)^3} \int dx_2 dp^2_{\perp} |\phi_{np}(x_2, p_{\perp})|^2 p_i p^{*}_j=\delta_{ij} \ ,
\label{normalP}
\eeq
where $p_i=(p^{+}, p^{-},p_z)$. In terms of the P-wave wavefunction in normal
coordinates,
\beq
\phi_{np} (x_2, p_{\perp})= 4\pi^{\frac{3}{2}} \sqrt{\frac{d p_z}{dx_2}}\psi_p (p) \ , \quad \frac{dp_z}{dx_2}= \frac{e'_1e_2}{x_1x_2 M'_{0}} \
\eeq
we have the following normalization condition:
\beq
\frac{1}{3} \cdot 4\pi  \int^{\infty}_{0}|\psi_{np}(p)|^2 p^4 dp = 1 \ .
\eeq

For the gaussian type 1P and 1S wavefunctions, we have the relation
\beq
\psi^{1P}_{p}(p)=\sqrt{\frac{2}{\beta^2}} \psi^{1S}_s (p) \ .
\eeq

The modified harmonic oscillator n-S wavefunctions in the light-front
approach are \cite{Ke:2010vn}
\beqs
\psi^{1S}_{s,M}(p) &=& \left( \frac{1}{\beta^2 \pi}  \right)^{\frac{3}{4}} \exp \left(  -\frac{1}{2}\frac{p^2}{\beta^2} \right) \nonumber \\
\psi^{2S}_{s,M}(p) &=& \left( \frac{1}{\beta^2 \pi}  \right)^{\frac{3}{4}} \exp \left(  -\frac{2^\delta}{2}\frac{p^2}{\beta^2} \right) \left( a'_2 -b'_2 \frac{p^2}{\beta^2} \right) \nonumber \\
 \psi^{3S}_{s,M}(p) &=& \left( \frac{1}{\beta^2 \pi}  \right)^{\frac{3}{4}} \exp \left(  -\frac{3^\delta}{2}\frac{p^2}{\beta^2} \right) \left( a'_3 -b'_3 \frac{p^2}{\beta^2} + c'_3
  \frac{p^4}{\beta^4} \right) \nonumber \\
\eeqs
where
\beqs
&& a'_2 = 1.88684 \quad b'_2 =1.54943   \nonumber \\
&& a'_3 = 2.53764 \quad b'_3 =5.67431 \quad c'_3 = 1.85652 \nonumber \\
&& \delta =1/1.82 \ .
\eeqs
The $n$P wavefunction, is related to $n$S wavefunction as follows:
\beq
\psi^{nP}_{p,M}(p)=\frac{C_n}{\beta}\psi^{nS}_{s,M}(p) \ ,
\eeq
where the constants $C_n$ can be determined by the normalization condition in
Eq.(\ref{normalP}) as
\beqs
&&C_1 =\sqrt{2}=1.41421\nonumber\\
&&C_2 =1.23833 \nonumber\\
&&C_3 =1.13215 \ .
\eeqs
%

\section{Some expressions in the light-front formalism}
\label{experssion}

In the covariant light-front formalism we have
\beqs
M'^2_0 &=&(e'_1 + e_2)^2 = \frac{p'^2_{\perp}+m'^2_1}{x_1}+ \frac{p'^2_{\perp}+m^2_2}{x_2} \nonumber\\
M''^2_0 &=&(e''_1 + e_2)^2 = \frac{p''^2_{\perp}+m''^2_1}{x_1}+ \frac{p''^2_{\perp}+m^2_2}{x_2}  \nonumber\\
\tilde M'_{0} &=&  \sqrt{M'^2_0-(m'_1-m_2)^2}\nonumber\\
\tilde M''_{0} &=& \sqrt{M''^2_0-(m''_1-m_2)^2} \nonumber\\
p'_z &=& \frac{x_2 M'_{0}}{2}- \frac{m^2_2+p'^2_{\perp}}{2x_2 M'_{0}} \nonumber\\
p''_z &=& \frac{x_2 M''_{0}}{2}- \frac{m^2_2+p''^2_{\perp}}{2x_2 M''_{0}} \nonumber\\
e'_1 &=&\sqrt{m'^2_1+p'^2_{\perp}+p'^2_{z}} \nonumber \\
e''_1 &=&\sqrt{m''^2_1+p''^2_{\perp}+p''^2_{z}} \nonumber \\
e_2 &=&\sqrt{m^2_2+p'^2_{\perp}+p'^2_{z}} \ ,
\eeqs

The explicit expressions for $A^{(i)}_{j} (i,j=1 \sim 4) $ are
\beqs
&&A^{(1)}_1=\frac{x_1}{2} \ , \ A^{(1)}_2= A^{(1)}_1- \frac{p'_{\perp}\cdot q_{\perp}}{q^2}\ , \nonumber\\
&&A^{(2)}_{1}=-p'^2_{\perp}-\frac{(p'_{\perp}\cdot q_{\perp})^2}{q^2}, \ A^{(2)}_2= (A^{(1)}_1)^2, \nonumber\\
&& A^{(2)}_3= A^{(1)}_1 A^{(1)}_2 \ , \   A^{(2)}_4= (A^{(1)}_2)^2-\frac{1}{q^2}A^{(2)}_1 \ ,\nonumber\\
&& A^{(3)}_1 = A^{(1)}_1 A^{(2)}_1 \ , A^{(3)}_2 = A^{(1)}_2 A^{(2)}_1 \ , \nonumber\\
&& A^{(3)}_3 = A^{(1)}_1 A^{(2)}_2 \ , A^{(3)}_4 = A^{(1)}_2 A^{(2)}_2 \  .
\eeqs
%


\end{appendix}


\newpage


\begin{thebibliography}{99}



\bibitem{Aubert:1974js}
  J.~J.~Aubert {\it et al.},
  Phys.\ Rev.\ Lett.\  {\bf 33}, 1404 (1974).

\bibitem{psi_slac}
J.~E.~Augustin {\it et al.},
  Phys.\ Rev.\ Lett.\  {\bf 33}, 1406 (1974).

\bibitem{Herb:1977ek}
  S.~W.~Herb {\it et al.},
  Phys.\ Rev.\ Lett.\  {\bf 39}, 252 (1977).

\bibitem{Innes:1977ae}
  W.~R.~Innes {\it et al.},
  Phys.\ Rev.\ Lett.\  {\bf 39}, 1240 (1977).


\bibitem{quiggrosner79}
C. Quigg and J. L. Rosner, Phys. Rept. {\bf 56}, 167 (1979).

\bibitem{grossemartin80}
H. Grosse and A. Martin, Phys. Rept. {\bf 60}, 341 (1980).

\bibitem{franzini82}
  P.~Franzini and J.~Lee-Franzini, Phys.\ Rept.\  {\bf 81}, 239 (1982).

\bibitem{kwong_rosner_quigg1987}
W. Kwong, J. L. Rosner, C. Quigg,
Ann. Rev. Nucl. Part. Sci. {\bf 37}, 325 (1987).

\bibitem{Brambilla:2004wf}
  N.~Brambilla {\it et al.} [Quarkonium Working Group Collaboration],
  hep-ph/0412158.

\bibitem{quarkonia2008}
E. Eichten, S. Godfrey, H. Mahlke, and J. L. Rosner, Rev. Mod. Phys.
{\bf 80}, 1161 (2008).

\bibitem{voloshin2008}
M. B. Voloshin, Prog. Part. Nucl. Phys. {\bf 61}, 455 (2008).

\bibitem{cesr}
K. Berkelman and E. H. Thorndike, Ann. Rev. Nucl. Part. Sci. {\bf 59},
297 (2009).

\bibitem{Brambilla:2010cs}
  N.~Brambilla {\it et al.},
  Eur.\ Phys.\ J.\ C {\bf 71}, 1534 (2011).

\bibitem{rosner2011}
J. L. Rosner, in Proc. of Ninth International Conference on Flavor Physics and
CP Violation (FPCP 2011, Israel), arXiv:1107.1273.

\bibitem{rosner2013}
C. Patrignani, T. K., and J. Rosner, Annu. Rev. Nucl. Part. Sci. {\bf 63}, 21
(2013).



\bibitem{Whitaker:1976hb}
  J.~S.~Whitaker {\it et al.},
  Phys.\ Rev.\ Lett.\  {\bf 37}, 1596 (1976).

\bibitem{Biddick:1977sv}
  C.~J.~Biddick {\it et al.},
  Phys.\ Rev.\ Lett.\  {\bf 38}, 1324 (1977).




\bibitem{klopfenstein83}
C. Klopfenstein et al., (CUSB Collab.), Phys. Rev. Lett. {\bf 51},
160 (1983).

\bibitem{pauss83}
F. Pauss et al. (CUSB Collab.), Phys. Lett. {\bf 130B}, 439 (1983).

\bibitem{haas84}
P. Haas et al. (CLEO Collab.), Phys. Rev. Lett. {\bf 52}, 799 (1984).

\bibitem{kornicer2011}
M. Kornicer et al. (CLEO Collab.), Phys. Rev. D {\bf 83}, 054003 (2011).

\bibitem{babar2014}
J. P. Lees et al. (BABAR Collab.), Phys. Rev. D {\bf 90}, 112010 (2014).

\bibitem{Aad:2011ih}
  G.~Aad {\it et al.} (ATLAS Collab.)
  Phys.\ Rev.\ Lett.\  {\bf 108}, 152001 (2012).


\bibitem{Aaij:2014hla}
  R.~Aaij {\it et al.} (LHCb Collab.)
  JHEP {\bf 1410}, 88 (2014).


\bibitem{Karl:1980wm}
  G.~Karl, S.~Meshkov and J.~L.~Rosner,
  Phys.\ Rev.\ Lett.\  {\bf 45}, 215 (1980).

\bibitem{Moxhay:1983vu}
  P.~Moxhay and J.~L.~Rosner,
  Phys.\ Rev.\ D {\bf 28}, 1132 (1983).

\bibitem{McClary:1983xw}
  R.~McClary and N.~Byers,
  Phys.\ Rev.\ D {\bf 28}, 1692 (1983).

\bibitem{Grotch:1984gf}
  H.~Grotch, D.~A.~Owen and K.~J.~Sebastian,
  Phys.\ Rev.\ D {\bf 30}, 1924 (1984).

\bibitem{Ebert:2002pp}
  D.~Ebert, R.~N.~Faustov and V.~O.~Galkin,
  Phys.\ Rev.\ D {\bf 67}, 014027 (2003).


\bibitem{Li:2009nr}
  B.~Q.~Li and K.~T.~Chao,
  Commun.\ Theor.\ Phys.\  {\bf 52}, 653 (2009).

\bibitem{Godfrey:2015dia}
  S.~Godfrey and K.~Moats,
  Phys.\ Rev.\ D {\bf 92}, 054034 (2015).


\bibitem{Segovia:2016xqb}
  J.~Segovia, P.~G.~Ortega, D.~R.~Entem and F.~Fernandez,
  Phys.\ Rev.\ D {\bf 93}, 074027 (2016).

\bibitem{Brambilla:2005zw}
  N.~Brambilla, Y.~Jia and A.~Vairo,
  Phys.\ Rev.\ D {\bf 73}, 054005 (2006)


\bibitem{Terentev:1976jk}
  M.~V.~Terentev,
  Sov.\ J.\ Nucl.\ Phys.\  {\bf 24}, 106 (1976)
  [Yad.\ Fiz.\  {\bf 24}, 207 (1976)].

\bibitem{Berestetsky:1976um}
  V.~B.~Berestetsky and M.~V.~Terentev,
  Sov.\ J.\ Nucl.\ Phys.\  {\bf 24}, 547 (1976)
  [Yad.\ Fiz.\  {\bf 24}, 1044 (1976)].


\bibitem{Lepage:1980fj}
  G.~P.~Lepage and S.~J.~Brodsky,
  Phys.\ Rev.\ D {\bf 22}, 2157 (1980).

\bibitem{Chung:1988mu}
  P.~L.~Chung, F.~Coester and W.~N.~Polyzou,
  Phys.\ Lett.\ B {\bf 205}, 545 (1988).

\bibitem{Brodsky:1997de}
  S.~J.~Brodsky, H.~C.~Pauli and S.~S.~Pinsky,
  Phys.\ Rept.\  {\bf 301}, 299 (1998)

\bibitem{brodsky_lf}
S. J. Brodsky and G. F. de Teramond, Phys. Rev. Lett. {\bf 96},
201601 (2006); G. F. de Teramond and S. J. Brodsky,
Phys. Rev. Lett. {\bf 102}, 081601 (2009);
S. J. Brodsky and G. F. de Teramond, Acta Phys. Polon. B {\bf 41},
2605 (2010).

\bibitem{brodsky_phys_rept}
S. J. Brodsky, G. F. de Teramond. H. G. Dosch, and J. Erlich,
Phys. Rept. {\bf 584}, 1 (2015).


\bibitem{Jaus:1989au}
  W.~Jaus,
  Phys.\ Rev.\ D {\bf 41}, 3394 (1990);
  W.~Jaus,
  Phys.\ Rev.\ D {\bf 44}, 2851 (1991).

\bibitem{Jaus:1999zv}
  W.~Jaus,
  Phys.\ Rev.\ D {\bf 60}, 054026 (1999).

\bibitem{Cheng:1996if}
  H.~Y.~Cheng, C.~Y.~Cheung and C.~W.~Hwang,
  Phys.\ Rev.\ D {\bf 55}, 1559 (1997).

\bibitem{Cheng:2003sm}
  H.~Y.~Cheng, C.~K.~Chua and C.~W.~Hwang,
  Phys.\ Rev.\ D {\bf 69}, 074025 (2004).

\bibitem{Hwang:2006cua}
  C.~W.~Hwang and Z.~T.~Wei,
  J.\ Phys.\ G {\bf 34}, 687 (2007)

\bibitem{Choi:2007se}
  H.~M.~Choi,
  Phys.\ Rev.\ D {\bf 75}, 073016 (2007).

\bibitem{Hwang:2010iq}
  C.~W.~Hwang and R.~S.~Guo,
  Phys.\ Rev.\ D {\bf 82}, 034021 (2010)

\bibitem{Ke:2010vn}
  H.~W.~Ke, X.~Q.~Li, Z.~T.~Wei and X.~Liu,
  Phys.\ Rev.\ D {\bf 82}, 034023 (2010).

\bibitem{Ke:2013zs}
  H.~W.~Ke, X.~Q.~Li and Y.~L.~Shi,
  Phys.\ Rev.\ D {\bf 87}, 054022 (2013).


\bibitem{Drell:1969km}
  S.~D.~Drell and T.~M.~Yan,
  Phys.\ Rev.\ Lett.\  {\bf 24}, 181 (1970).



\bibitem{Dudek:2006ej}
  J.~J.~Dudek, R.~G.~Edwards and D.~G.~Richards,
  Phys.\ Rev.\ D {\bf 73}, 074507 (2006)

\bibitem{Cho:1994gb}
  P.~L.~Cho, M.~B.~Wise and S.~P.~Trivedi,
  Phys.\ Rev.\ D {\bf 51}, R2039 (1995)

\bibitem{Shao:2012fs}
  H.~S.~Shao and K.~T.~Chao,
  Phys.\ Rev.\ D {\bf 90}, 014002 (2014)


\bibitem{Cheung:1995ub}
  C.~Y.~Cheung, W.~M.~Zhang and G.~L.~Lin,
  Phys.\ Rev.\ D {\bf 52}, 2915 (1995)

\bibitem{Isgur:1988gb}
  N.~Isgur, D.~Scora, B.~Grinstein and M.~B.~Wise,
  Phys.\ Rev.\ D {\bf 39}, 799 (1989).

\bibitem{Choi:1997iq}
  H.~M.~Choi and C.~R.~Ji,
  Phys.\ Rev.\ D {\bf 59}, 074015 (1999).

\bibitem{Choi:1999nu}
  H.~M.~Choi and C.~R.~Ji,
  Phys.\ Lett.\ B {\bf 460}, 461 (1999).


\bibitem{Eichten:1978tg}
  E.~Eichten, K.~Gottfried, T.~Kinoshita, K.~D.~Lane and T.~M.~Yan,
  Phys.\ Rev.\ D {\bf 17}, 3090 (1978)
  Erratum: [Phys.\ Rev.\ D {\bf 21}, 313 (1980)].

\bibitem{Eichten:1979ms}
  E.~Eichten, K.~Gottfried, T.~Kinoshita, K.~D.~Lane and T.~M.~Yan,
  Phys.\ Rev.\ D {\bf 21}, 203 (1980).

\bibitem{Eichten:1976jk}
  E.~Eichten and K.~Gottfried,
  Phys.\ Lett.\ B {\bf 66}, 286 (1977).


\bibitem{Eichten:1994gt}
  E.~J.~Eichten and C.~Quigg,
  Phys.\ Rev.\ D {\bf 49}, 5845 (1994).

\bibitem{Buchmuller:1980su}
  W.~Buchmuller and S.~H.~H.~Tye,
  Phys.\ Rev.\ D {\bf 24}, 132 (1981).

\bibitem{Gupta:1982kp}
  S.~N.~Gupta, S.~F.~Radford and W.~W.~Repko,
  Phys.\ Rev.\ D {\bf 26}, 3305 (1982);
  S.~N.~Gupta, S.~F.~Radford and W.~W.~Repko,
  Phys.\ Rev.\ D {\bf 34}, 201 (1986).

\bibitem{Kwong:1988ae}
  W.~Kwong and J.~L.~Rosner,
  Phys.\ Rev.\ D {\bf 38}, 279 (1988).


\bibitem{Godfrey:1985xj}
  S.~Godfrey and N.~Isgur,
  Phys.\ Rev.\ D {\bf 32}, 189 (1985).

\bibitem{Laermann:1986pu}
  E.~Laermann, F.~Langhammer, I.~Schmitt and P.~M.~Zerwas,
  Phys.\ Lett.\ B {\bf 173}, 437 (1986).

\bibitem{Chao:1992et}
  K.~T.~Chao, Y.~B.~Ding and D.~H.~Qin,
  Commun.\ Theor.\ Phys.\  {\bf 18}, 321 (1992).

\bibitem{Ding:1993uy}
  Y.~B.~Ding, K.~T.~Chao and D.~H.~Qin,
  Chin.\ Phys.\ Lett.\  {\bf 10}, 460 (1993).

\bibitem{Sumino:2001eh}
  Y.~Sumino,
  Phys.\ Rev.\ D {\bf 65}, 054003 (2002).

\bibitem{Recksiegel:2001xq}
  S.~Recksiegel and Y.~Sumino,
  Phys.\ Rev.\ D {\bf 65}, 054018 (2002).


\bibitem{PDG}
C. Patrignani et al. (Particle Data Group), Chin. Phys. C, 40, 100001 (2016);
online at http://pdg.lbl.gov.


\end{thebibliography}
\end{document}